\documentclass[preprint,showpacs,preprintnumbers,amsmath,amssymb]{revtex4}

\usepackage{graphicx}
\usepackage{dcolumn}
\usepackage{bm}
\usepackage{cancel}
\usepackage{amsmath}
\def\rvarphi{\raise 2pt\hbox{$\varphi$}}
\def\reta{\raise 1pt\hbox{$\eta$}}
\DeclareMathAlphabet{\mathscr}{OT1}{pzc}{m}{it}

\begin{document}


\title{A Spinor Approach to Penrose Inequality}

\author{Yun-Kau Lau}
 \affiliation {Institute of Applied Mathematics, Morningside Center of Mathematics, Academy of
Mathematics and System Science, Chinese Academy of Sciences, 55,
Zhongguancun Donglu, Beijing, 100190, China. }

\vskip 80pt

\begin{abstract}
Consider an asymptotically Euclidean initial data set  with a smooth marginally
trapped surface (possibly a union of future and past multi-connected
components) as inner boundary.  By a further development of the spinorial framework underlying the positive energy theorem, a refined Witten identity is worked out and in the maximal slicing case, a close connection of the identity with a conformal invariant of Yamabe type is revealed. A Kato-Yau inequality for the Sen-Witten operator is also proven from
a conformal geometry perspective. Guided by the
Hamiltonian picture underlying  the spinorial framework, a Penrose type inequality is then proven to the effect
that given the dominant energy condition,  the   ADM energy-momentum is, up to a  non-zero constant less than unity, bounded by the
areal radius of the marginally trapped surface.
To establish the Penrose inequality in full generality, it  is then  sufficient to
   show that the  norm of
 the Sen-Witten spinor, subject to the APS boundary condition imposed
  on a suitably defined outermost marginally trapped surface, is bounded below by that attained
 in the Schwarzschild metric.
\end{abstract}


\maketitle

\section{Introducton}

When the Penrose inequality is regarded
as a strengthened form of the positive energy theorem for black holes,
it is natural to ask whether the spinorial proof of the positive
energy theorem, first initated by Witten \cite{witten}, may be
suitably generalised to tackle the Penrose inequality,
particularly in the outstanding case when the initial data set is not time symmetric (see  \cite{bray,huisken} in the time
symmetric case).

Given the three manifold of an initial data set, underlying the
spinorial approach  to the positive energy theorem is the  physical picture that a
non-zero spinor field together with its dual (defined in terms of the
timelike unit normal of the three manifold in spacetime) generate a Newman-Penrose tetrad,
 from which  an orthonormal moving frame is
further defined and plays the role of canonical variables in
describing the Hamiltonian dynamics of a gravitational field
\cite{AH,nester}. In this sense, the Sen-Witten equation may be
regarded as a gauge condition to select a moving frame on a three manifold
  (see also \cite{parker}) to parametrise the
Hamiltonian.

To explore a spinorial approach to the Penrose
inequality,  so far two
obstacles have been encountered. The first one is the need to
further develop
Witten's spinorial technique by taking the fourth root of it, in a sense to be made precise in what follows. Another obstacle is the incompatibility of the APS boundary condition imposed on a spinor field with the marginally trapped boundary condition imposed on the inner boundary.
 The flagpole of the Sen-Witten spinor field subject to the APS boundary condition in general will not align with one of the two null normals of the marginally trapped surface under consideration.
  We
   shall
seek to address these two issues in the present work and it
turns out a better understanding of the Hamiltonian picture
underlying the spinorial approach enables us to find a way to go forward.
 A Penrose type inequality involving the ADM energy-momentum for a
generic asymptotically Euclidean initial data set then
emerges naturally for the first time. The obstacle to a complete proof of the Penrose inequality is also identified.

For a good description of the Penrose inequality, see \cite{unsolved}.
 A review
of the  the Penrose inequality may be
found in \cite{mars}. To tackle the Penrose
inequality using spinors was also considered in \cite{herzlich} and further generalised in \cite{khuri}, with however only the ADM mass
considered.
As we shall see in what follows, the line of argument presented here is Hamiltonian in essence  and in many ways distinct from the previous
spinor approach.

The outline of the article may be given as follows. After certain
preliminaries in Section 2,  in Section 3 we shall seek to further develop the spinorial framework used in the proof of the positive energy theorem
and a new refined Witten identity is worked out. Certain geometric structures underneath the refined identity will also be discussed.
By twisting the Sen-Witten spinor field in
a sense to be described, a new shift vector for the
Hamiltonian is defined in Section 4 and its obstruction to the positivity of the refined Witten identity is addressed.  A Penrose type inequality  for the ADM
energy momentum for a generic asymptotically Euclidean initial data
set is then presented for the first time.  The rest of the paper then serves to fill in the
details of the proof of the main theorem presented in Section 4, including the derivation of a refined Witten identity, regularisation of zero points of the Sen-Witten spinor field and the proof of existence and uniqueness of the Sen-Witten
spinor field, given the APS boundary condition at the inner boundary and appropriate falloff near spatial infinity.

\section{Preliminaries and notations}

Some background materials  relevant to the present work will be
briefly described in this section.  The notations for two spinors
will follow that in \cite{pr} unless otherwise stated.

 Let $(M,g_{ab})$ be a
 smooth, connected four dimensional  spacetime manifold with Lorentzian
 metric signature
   $(+,\,-,\,-,\,-)$. Suppose $N$ is an orientable, complete
Riemannian three manifold  identically embedded in $M$  so that when
restricted to $N$,
\begin{equation*}g_{ab}=
\tau_a\,\tau_b-h_{ab},\end{equation*} where   $h_{ab}$ is a smooth
Riemannian metric of $N$ and $\tau^a$ is the unit timelike normal of
$N$ in $M$. $N$ is assumed to be asymptotically Euclidean in the
standard sense that in
 the complement of some compact set
in $N$, \begin{equation*}h_{ab}=\eta_{ab}+\,O(1/r),\end{equation*}
$\eta_{ab}$ is an Euclidean metric and \begin{equation*}\partial
h_{ab}=O(1/r^{2}),
 \quad \partial^2 h_{ab}=O(1/r^{3}),
\end{equation*}
where  $r$ is the standard radial parameter defined in terms of the
Cartesian coordinates near infinity. When $N$ is considered as a
spacelike hypersurface identically embedded in $M$,  the second
fundamental form of $N$ in $M$ is given by
$K_{ab}=h_a^lh_b^m\nabla_l \tau_m $ and in the asymptotic regime,
$$K_{ab}=O(1/r^{2}), \quad
\partial\, K_{ab}=O(1/r^{3}).$$

As a codimension one submanifold of  $(M,g_{ab})$, the geometry of
$(N, h_{ab},K_{ab})$ is also subject to the Hamiltonian and momentum
constraint  equations given respectively by
\begin{eqnarray}
R&=&2\mu\,+\,|K_{ab}|^2\,-\,K^2,\label{hcon}\\
j_b&=&D^a (K_{ab}-Kh_{ab})\label{mcon},
\end{eqnarray}
where $R$ is the scalar curvature of $(N,h_{ab})$,  $K=h^{ab}K_{ab}$,
  $\mu$ and $j_a$ are respectively
 the  density and current of local matter
as measured by an observer at rest with respect to $N$. The four
vector $(\mu,j^a)$ is required to satisfy the dominant energy condition
$\mu\,\ge j^aj_a$ throughout the present work.

Denote by $\partial N$ the inner boundary of $N$. $\partial N$ is
assumed to  consist of connected components $S_i$, $i=0,1,\cdots n$
with each $S_i$  a smooth spherical two surface. Let $\gamma_{ab}$
and $p\,$ be respectively the two metric  and
 the mean curvature  of $S_i$ defined with respect to the outward pointing normal.
  Then
\begin{equation}
\label{trapped}
  \gamma^{ab}K_{ab}\pm p=0
  \end{equation}
   characterise $S_i$ as a future $(+)$ and past
  $( -  )$
   marginally trapped surface.

 Denote by $\tau^{AA'}$  the
timelike unit normal of $N$ in spinorial indices.
Let $\nabla_{AA'}$ be the spin connection lifted from the metric
connection of $(M,g_{ab})$, the projection of $\nabla_{AA'}$ on $N$
may be given as \cite{sen}
  \begin{equation}
  \label{sw00}{\mathscr{D}}_{\,\,AB}:=\sqrt
2\, \tau_{(B}{}^{A'}\nabla_{A)A'}.\end{equation}
Denote by ${{D}}_{AB}$ the spin connection of $(N, h_{ab})$, it  may be defined in terms of ${\mathscr{D}}_{\,\,AB}$ as
\begin{equation}
\label{swop}D_{AB }\lambda _C={\mathscr
D}_{AB}\lambda_C\,-\,\frac{\,1\,}{\,\sqrt
2\,}K_{ABCD}\lambda^D,\end{equation}
 where $K_{ABCD}=2\,
 \tau_{B}{}^{A'}\tau_{D}{}^{C'}K_{AA'CC'}$ and
 $K_{AA'CC'}$ is the second fundamental form of  $N$ in spinorial indices.

  We shall adopt the
following Sen-Witten equation as the gauge condition to specify a spin frame in $(N, h_{ab}, K_{ab})$ given by
 \begin{equation}
\label{sw}
  {\mathscr D}_{A}{}^{C}\lambda_{\,C}=0.
\end{equation}
 Away from the zero points of $\lambda^A$,
a non-trivial dual of $\lambda^A$ may be defined in terms of $\tau_{AA'}$ as
\begin{equation*}\lambda ^{\dagger A}=\sqrt{2}\, \tau^{AA'}\lambda
_{A'}.\end{equation*}
 We further subject $\lambda^A$ to the
 asymptotic boundary conditions that, near
infinity,
\begin{equation*}
\label{asy-boundary}
 \lambda^A=\lambda_{\,0}^A+O(1/r)
 \end{equation*}
  where
  $\lambda_{\,0}^A$ is a covariantly constant spinor defined with
respect to the flat connection of $\eta_{ab}$. At the inner boundary
$S$, let $\cancel\nabla_{AC}$ be the spin connection pertained to the two metric of $S$.
$\lambda^A$ is said to satisfy the APS (spectral) boundary condition at $S$
\cite{aps} (see also \cite{herzlich}) in that
\begin{equation}
\label{aps1}
 \lambda_A=\sum_{n=0}^{\infty}\, a_n\lambda_{nA},\quad
a_n\in C.
\end{equation}
$\lambda_{nA}$ are eigenspinors given by
\begin{equation*}
\label{aps2} \cancel\nabla_{A}{}^C\lambda_{nC}=-\frac{\,1}{\sqrt
2}\mu_n\lambda_{nA}, \mu_n>0 \quad\hbox {for}\,\, n=0,1,2\dots
\end{equation*}
and $\{\lambda_n^A\}_{n=0,1,2...}$ constitute an orthonormal basis
defined by the natural $l^2$ scalar product. $|\phantom{a} |^2$ denotes the hermitan norm of a spinor field defined with respect to $ \tau^{AA'}$.

Throughout the present work, contraction of
tensorial and spinorial
indices are always defined with respect to $h_{ab}$ and the
symplectic form $\epsilon_{AB}$ respectively unless otherwise stated.

\section{Development of the spinorial framework}

Let us begin by looking at the simple example of a constant time slice of the Schwarzschild metric, whose metric is given by
 \begin{equation*}
 ds^2=\Big(1+\frac{M}{2r}\Big)^4(dr^2+r^2d\Omega^2).
 \end{equation*}
 Calculations on this simple example suggest that the conventional spinorial approach will not yield an optimal
 Penrose inequality. Instead, we need to further develop the Witten identity by taking its fourth root  in the following sense.

Define
\begin{equation}
\label{fourth}
 u^4=\lambda_A\lambda^{\dag A}.
 \end{equation}
 Provisionally we assume $u>0$ (i.e. $\lambda^A$ is non-zero everywhere in $N$)  and seek to relax
  this later. The example of the
Schwarzschild metric leads us to adopt the following definition of a
two surface functional.
 \begin{equation}
\label{quasi} M(S)=\frac{1}{2\pi}\int_{S}\, D_{a}u\,dS^a,
\end{equation}
where $S$ is a spherical two surface embedded in $N$.
 For a round
sphere of radius $r\ge \frac{M}{2}$ in a constant time slice of the
Schwarzschild metric, (\ref{quasi}) always yields $M$. So at least
in this simple example, the definition in (\ref{quasi}) resembles
the Hawking mass in that it
 yields the irreducible mass for a black hole at the
outermost marginally trapped surface and the ADM mass at infinity.

A couple of remarks (caveats) of the definition are in order here.
In the simple case of Euclidean $R^3$ with a
non-round sphere chosen as the inner boundary, the mass functional yields negative value
 and goes to zero at infinity from below zero. This turns out to be a blessing in disguise and is related
 to a more general Minkowski inequality in Euclidean $R^3$. This  problem will be taken up elsewhere. In the present context,
  we shall take a pragmatic stand
 and look on the definition as a useful handle to linking up
the ADM energy at spatial infinity and a spinorial analog at a
marginally trapped surface.  Further, at points
where
$\lambda^A$ is zero, pointwise the gradient term $D_a u$ becomes
singular. We will address this problem later on.

Given $u$ defined in (\ref{fourth}), the next natural step to take is to work out a Witten type identity  for it. Written in terms of $u$, the Hamiltonian part of the conventional Witten identity may be given as
\begin{eqnarray}
\triangle u^4&=& 4u^3\triangle u+4u^2D_auD^au\nonumber\\
&=&(\mu+\frac{1}{2}|K_{ab}|^2)u^4+2\,|{D}_{AB}\lambda_C|^2-\sqrt 2\lambda^A\lambda^{\dag B}D_{AB}K\label{swii0}
 \end{eqnarray}
By our provisional hypothesis,
 $u>0$, we may normalise
$\lambda_A,\lambda_A^\dagger$ and define a spin frame
 $(o_A,\iota_A)$ by
\begin{equation} \label{spin}
\lambda_A=u^2o_A,\quad\lambda_A=u^2\iota_A.
\end{equation}
 (\ref{swii0}) may then be written as
\begin{eqnarray}
\label{swiii}
 &&4u^3\triangle u+4u^2D_auD^au\nonumber\\
 &=&(\mu+\frac{1}{2}|K_{ab}|^2)u^4+2u^4\,|{D}_{AB}o_C|^2-\sqrt 2 u^4o^A\iota^B D_{AB}K.
\end{eqnarray}

To elaborate (\ref{swiii}) further, we shall  exploit the conformal rescaling symmetries of the Sen-Witten equation.
Define
\begin{equation}
 \label{scal2}
\hat \tau_a=\,u^2\tau_a,\,\, \hat h_{ab}=u^4h_{ab}.
\end{equation}
In the simple case of the Schwarzschild metric, conformal flatness means that $\hat h_{ab}$ is just the Euclidean metric.
Denote by $\hat{\mathscr D}_{AB}$ the conformally rescaled Sen-Witten connection defined in terms of $\hat\tau_a$ and $\hat h_{ab}$ given above.
Conformal rescaling symmetry of (\ref{sw}) means that we also have
\begin{equation}
\label{dir}
 \hat\epsilon^{BC}\hat {{\mathscr D}}_{AB}\hat\lambda_C=0
\end{equation}
with
\begin{equation}
\label{11}\hat\lambda_C=u^{-1}\lambda_C=u\,o_C, \quad
\hat\lambda_C^\dagger=u^{-1}\lambda^\dagger _C=u\,\iota_C
\end{equation}
according to (\ref{spin}) and $\hat\epsilon^{AM}=u^{-2}\epsilon^{AM}$ is the conformally rescaled symplectic form.
 It may further be checked that
$\hat\epsilon^{AB}\hat\lambda_A\hat\lambda^\dagger_B=1$
and therefore
$(\hat\lambda_C, \hat\lambda_C^\dagger)$ generate a spin frame under
 $\hat\epsilon^{AB}$.
 Using the Sen-Witten equation and after some very tedious spinor calculus, we work out the following spinor identity
\begin{eqnarray}
 \phantom{u^4}|{  D}_{AB}o_C|^2
 =
 u^4|\hat { D}_{AB} \hat \lambda_C|^2
+2\,|D_a\ln u|^2-K\nu^aD_a\ln u\label{finn}
\end{eqnarray}
where $\hat {D}_{AB}$ is the conformally rescaled spin connection of $ { D}_{AB}$ and $\nu_a=\sqrt 2o_{(A}\iota_{B)}$. Details of the derivation of (\ref{finn}) will be presented later on. Let us check that in the maximal slicing case when $K=0$, we may infer from (\ref{finn}) the following Kato-Yau inequality for  a harmonic spinor field expressed as
\begin{equation} \label{KY2}
 |{ D}_{AB}\lambda_C|^2\ge\,\frac{3}{2} |D_a \left|\lambda\,|\right|^2
\end{equation}
where $|\lambda|=u^2$ (cf~\cite{KY} and references therein).  This may be regarded as a consistency check on the validity of the spinor identity in (\ref{finn}) and at the same time gives a new proof of the Kato-Yau inequality for harmonic spinor field from a conformal geometry perspective.

 Given (\ref{finn}), (\ref{swiii}) may be further expressed as
 \begin{eqnarray}
\label{swiiii}
 \triangle u&=&\frac{1}{4}(\mu+\frac{1}{2}|K_{ab}|^2)u+\frac{1}{2}u^5|\hat { D}_{AB}\hat \lambda_C|^2\nonumber\\
 &&+\frac{1}{4}u\nu^aD_aK-\frac{1}{2} K\nu^a D_a u.
\end{eqnarray}
With the momentum constraint further taken into account and the shift vector chosen to be $N_a=u\nu_a$, it follows from (\ref{swiii}) and (\ref{finn}) that
\begin{eqnarray}
&&\triangle u \,-\,\frac{1}{4}
D^a(K_{ab}N^b)\nonumber\\
&=&\frac{1}{4}(\mu \,-\,j^a\nu_a)u\,+ \,\frac{1}{2}u^5|\hat { D}_{AB}\hat \lambda_C|^2\nonumber\\
&&+\frac{1}{4}\big(\,\frac{1}{2}\,|K_{ab}|^2-K_{ab}D^a\nu^b\big)u+\frac{1}{4}K_{ab}\nu^aD^bu- \frac{1}{2}K\nu^a D_a u\nonumber\\
&=&\frac{1}{4}(\mu\,-\,j^a\nu_a)u\,+ \,\frac{1}{2}u^5|\hat {D}_{AB}\hat \lambda_C|^2\nonumber\\
&&+\frac{1}{4}\big(\,\frac{1}{2}\,|K_{ab}|^2-K_{ab}D^a\nu^b\big)u-\frac{1}{2} K\nu^a D_a u
\label{refined-00}
 \end{eqnarray}
 where $|\hat {D}_{AB}\hat \lambda_C|$ is defined in terms of the conformally rescaled symplectic form $\hat\epsilon_{AB}$.
  In terms of the definition of $K_{ab}$ and some simple spinor calculus, it may be worked out
 that the spurious term $K_{ab}\nu^aD^bu$ vanishes in the first equality in (\ref{refined-00}).
 Subject to the conformal rescaling given in (\ref{scal2}), we have
 \begin{equation}
 \label{conformal-k}
 \hat K_{ab}=u^2K_{ab},\quad \hat D_{(a}\hat\nu_{b)}=u^2D_{(a}\nu_{b)}-2h_{ab}\nu^aD_au
 \end{equation}
 where $\hat\nu_a=u^2\nu_a$.
 From (\ref{conformal-k}), it may be deduced that
 \begin{equation}
 \label{conformal-l}
 \frac{1}{2}\,|K_{ab}|^2-K_{ab}D^a\nu^b=\,\big(\,\frac{1}{2}\,|\hat K_{ab}|^2-\hat K_{ab}\hat D^a\hat\nu^b\big)\,u^4
 \end{equation}
 where contraction of indices on the right hand side of (\ref{conformal-l}) is defined in terms of $\hat h_{ab}$.
 With (\ref{conformal-l}) input into (\ref{refined-00}), we then find
 \begin{eqnarray}
&&\triangle u \,-\,\frac{1}{4}
D^a(K_{ab}N^b)\nonumber\\
&=&\frac{1}{4}(\mu\,-\,j^a\nu_a)u\,+ \,\frac{1}{2}u^5|\hat {\mathscr D}_{AB}\hat \lambda_C|^2
\label{refined}
 \end{eqnarray}
 which may be regarded as  a refinement of the conventional Witten identity, with the fourth root of the spinor norm $u$ in place of the spinor norm $\varphi$ in the identity.

From  (\ref{finn}) together with the definition of the Sen-Witten operator in (\ref{sw00}),  a Kato-Yau inequality for the Sen-Witten operator  may also be worked out for the first time to be
\begin{equation} \label{KY-sw}
 |\mathscr{ D}_{AB}\lambda_C|^2\ge\,\frac{3}{2} |D_a \left|\lambda\,|\right|^2.
\end{equation}

Further, in the maximal slicing case, the Hamiltonian part of the refined Witten identity in (\ref{refined}) gives
\begin{equation}
\label{refined-maximal}
 \triangle u=\frac{1}{8}Ru+\frac{1}{2}u^5|\hat D_{AB}
\hat \lambda_C|^2.
\end{equation}
(\ref{refined-maximal}) resembles a conformal Laplacian
 if we formally identify the scalar curvature defined by the metric connection of  $\hat h_{ab}$ as
 $
  \hat R=-4\,|\hat D_{AB} \hat \lambda_C|^2.
  $
 This formal identification  actually gains weight if we work out
  the Witten identity for $\hat\lambda_A$.

The resemblance of (\ref{refined-maximal}) to a conformal Laplacian
leads us to consider the following conformal invariant appearing
 naturally in the Yamabe problem. For a real valued function $f$
in $N$, consider the following functional
\begin{equation}
\label{yamabe}
\int_{N}\,|D_af|^2+\frac{1}{8}\,R\,f^2\,+\,\frac{1}{4}\Big(\int_{S_\infty}f^2\,p-\,\int_{S}f^2\,p
\,\Big),
\end{equation}
 where $|D_af| ^2=h^{ab}D_a fD_b f$, $p$ is the mean
curvature of the boundary $S_\infty\cup S$ with the normal
of the boundary outward pointing. $S_\infty$ is a coordinate sphere
near spatial infinity while $S$ is the inner boundary.
 Instead of the standard choice of compactly supported test
functions, we allow $f$ to behave asymptotically as $f=f_0+O(1/r)$
for some constant $f_0$.

 As the choice of test functions in  the functional
(\ref{yamabe}) is no longer restricted to be compactly supported and
allowed to be asymptotically constant, we may choose $u^2$ as a test
function and the functional in (\ref{yamabe}) becomes
\begin{eqnarray}
&&\int_{N}\,|Du^2|^2+\frac{1}{8}\,R\,u^4\,+\,\frac{1}{4}\Big(\int_{S_\infty}u^4\,p-\,\int_{S}u^4\,p\,\Big)\nonumber\\
&=&\int_{\hat N}\,|\hat Du|^2\,+\,\frac{1}{8}\,\hat R\,
u^2\,+\,\frac{1}{4}\Big(\int_{S_\infty}\,u^2\hat p\,- \int_{S}\,u^2\hat p\,\Big).\label{yam2}
\end{eqnarray}
with
\begin{equation*}
u^4\hat p=u^{2}p\,+\, 2\nu^aD_a u^2, \quad \hat R=-4\,|\hat D_{AB} \hat \lambda_C|^2.
\end{equation*}
By rearranging terms in (\ref{yam2}), we find
\begin{eqnarray}
8\pi M\,-\,\int_{S}D_a u^4\,dS^a
&=&\int_{N}\,\,\,\frac{1}{2}\,R\,u^4\,+\,2| D_{AB}\lambda_C|^2
\label{yam3}
\end{eqnarray}
where $M$ is the ADM mass and we recover  the  conventional Witten identity in integral form. When the test
function is chosen to be $u^{\frac{\,1}{\,2}}$, we have
\begin{eqnarray}
2\pi M-\int_{S}D_a u\,dS^a
=\int_{N}\,\,\,\frac{1}{8}\,R\,u\,+\,\frac{1}{2}u^5 |\hat
D_{AB}\hat\lambda_C|^2  \label{yam4}
\end{eqnarray}
and this is just the Hamiltonian part of the refined Witten identity given in
(\ref{refined-maximal}) in integral form when $K=0$.
 In the maximal slicing case, both the Witten identity and its refined
 version in integral form
are merely a rearrangement of the terms in the conformal
invariant displayed in (\ref{yamabe}).

\section{ Twisted Sen-Witten spinor field.}

Unlike in the case of positive energy theorem, the refined Witten identity in (\ref{refined}) cannot be
applied in a straightforward manner to generate a Penrose type inequality. Calculations of some simple examples suggest that, subject to the APS boundary condition on $\lambda^A$, the  flagpole of
 $\lambda^{A}$ in general will not align with the null normals of $S$.
 This mismatch becomes a problem when we try to realise the
marginally trapped boundary condition in terms of $\lambda^A$.

To overcome this obstacle, bear in mind that
the choice of lapse and shift for a Hamiltonian is by no means
unique. Consideration of the time symmetric case  suggests that the fourth root of the spinor norm $u$ defined  by the Sen-Witten equation remains a good choice for the lapse function. However, from a physical standpoint,  a shift vector is not necessarily dictated by
the flagpole of $\lambda^A$ as in the proof of the positive energy theorem. What we will do is to  twist
$\lambda^A$  near $S$  by the standard cut and paste
technique in   such a way to force the flagpole of the twisted Sen-Witten spinor to align with one of the null normals of $S$.
Yet at the same time, the Sen-Witten equation satisfied by $\lambda^A$ is not disturbed.

 To proceed, compactness of the inner boundary $S$ enables us to infer the existence of some sufficiently small $\delta>0$ (to be kept fixed hereafter) such that
  near $S$ there exists a smooth one parameter family
of two spheres $S_x$ with $x\in[0,\delta]$. Let $N_{\epsilon}=\cup\, S_x, x\in[0,\epsilon) $, $\epsilon<\delta $ and denote by $(\,\tilde o^A, \tilde \iota^A)$
a spin frame with the two null normals of $S$ as flagpoles. Parallel
transport of $(\,\tilde o^A, \tilde \iota^A)$ along the affinely
parametrsied geodesics orthogonal to $S$ generates in $N_\epsilon $
two linearly independent spinor fields again denoted by $ (\,\tilde
o^A, \tilde \iota^A)$.

Introduce a cutoff function $\eta:N\rightarrow R$  such that
\begin{equation}
\label{cutoff1}
 0\le\eta\le 1, \quad |D_a\eta|\le 1
\end{equation}
in $N_\epsilon$ and zero elsewhere in $N$.
   Define a twisted spinor field
$\alpha_A$ in $N$ as
\begin{equation}
\label{alpha} {\alpha}_A=u^{\frac{1}{2}}(\eta\,\tilde{o}_A\,+\,(1-\eta)\,{o }_A)
\end{equation}
so that at $S$ the flagpole of ${\alpha}_A$ aligns with the null normal of $S$ defined by $\tilde o_A$ and in
$N/ N_\epsilon$, up to a scaling factor ${\alpha}_A$ agrees with the Sen-Witten spinor field.
In terms of $\alpha_A$, a shift vector of the Hamiltonian
may then be defined as
\begin{equation}
\label{shift-alp} n_a=\sqrt
2\,\,\alpha_{(A}\alpha^{\dag}{}_{B)}.
\end{equation}
It may be checked, using
 (\ref{cutoff1}) and (\ref{alpha})  that,
\begin{equation}
\label{shift-alp-0}
\alpha_a\alpha^{a}\le 1
\end{equation}
and therefore
 the four
vector $(u, n_a)$ is non-spacelike, as required by the
non-spacelike Hamiltonian evolution of the initial data set $(N, \,h_{ab},\,K_{ab})$.

When the shift vector is no longer dictated by the flagpole of the Sen-Witten spinor field $\lambda^A$, for an arbitrary shift vector $n^a$, the refined Witten identity in (\ref{refined}) may be written in a more general form as
\begin{eqnarray}
&&\triangle u\,-\,\frac{1}{4}D^a(K_{ab}n^b)\nonumber\\
&=&\frac{1}{4}(\mu u-j^an_a)+\frac{1}{2}u^5\,|\hat D_{AB}\hat\lambda_C|^2+\frac{1}{4}\,\Big(\,\frac{1}{2}|K_{ab}|^2u-K^{ab}D_an_b\Big)\nonumber\\
&& +\frac{1}{4}u(\nu^a-n^a)D_aK.\label{refined-k}
\end{eqnarray}
Given the lapse and shift specified respectively by $u$ and
$n_a$,
\begin{equation}
\label{kab}
K_{ab}=-\frac{1}{2u}(\dot{h}_{ab}-D_an_b-D_bn_a)
\end{equation}
where $\dot{h}_{ab}$ denotes the Lie derivative of $h_{ab}$ with
respect to the timelike vector field generating the Hamiltonian
evolution of $N$, it  follows  from (\ref{kab}) that
\begin{equation}
\label{kab0}
\frac{1}{2}|K_{ab}|^2u-K^{ab}D_an_b=\frac{1}{8u}\,|\dot{h}_{ab}|^2-\frac{1}{2u}|D_{(a}
n_{b)}|^2.
\end{equation}
Putting (\ref{kab0}) back into (\ref{refined-k}), we  have
\begin{eqnarray}
&&\triangle u\,-\,\frac{1}{4}D^a(K_{ab}n^b)\nonumber\\
&=&\frac{1}{4}(\mu u-j^an_a)+\frac{1}{2}u^5\,|\hat D_{AB}\hat\lambda_C|^2+\frac{1}{4}\Big(\,\frac{1}{8u}\,|\dot
h_{ab}|^2-\frac{1}{2u}|D_{(a}n_{b)}|^2\Big)\nonumber\\
&& +\frac{1}{4}u(\nu^a-n^a)D_aK.\label{1}
\end{eqnarray}
 By construction, the vector $(u,n_a)$ is non-spacelike and in view of the dominant energy condition, we may see that the obstruction to positivity comes from the  terms $|D_{(a}n_{b)}|^2$ and $(\nu^a-n^a)D_aK$ in the above expression.

Let $M-|P|$ be the Minkowski norm of the ADM energy-momentum four
vector at spatial infinity. By integrating (\ref{1}) over a region of $N=N_\epsilon\cup  N/N_\epsilon$ bounded by the inner boundary $S$ and a limiting
coordinate sphere $S_\infty$ at  infinity,  we have
\begin{eqnarray}
&&2\pi(M-|P|)\nonumber\\
&\ge&\,\int_{N_\epsilon\cup  N/N_\epsilon}\Big[\frac{1}{4}(\mu u-j^an_a)+\frac{1}{2}u^5\,|\hat D_{AB}\hat\lambda_C|^2\nonumber\\
&&+\frac{1}{4}\Big(\,\frac{1}{8u}\,|\dot
h_{ab}|^2-\frac{1}{2u}|D_{(a}n_{b)}|^2\Big) +\frac{1}{4}u(\nu^a-n^a)D_aK.\Big]\nonumber\\
&&+\frac{1}{4}\int_{S}\,\,-\sqrt
2\,u^{-3}(\lambda^{\dag
A}\,\cancel\nabla_{A}{}^C\lambda_C\,+\,\lambda_{A}\,\cancel\nabla^{AC}\lambda_C^{\dag})\,\nonumber\\
&&-(Ku-K_{ab}n^{b}\nu^a+p\,u).
\label{final2}
\end{eqnarray}
From (\ref{shift-alp-0}), we see that    $n^a=u\,\tilde\nu^a$ at $S$ where $\tilde\nu^a$ is the
outward pointing normal of $S$. It then follows from the  marginally trapped condition given in
(\ref{trapped}) that the curvature term in the inner boundary integral in (\ref{final2}) vanishes.
Further, by (\ref{alpha}) and (\ref{shift-alp}), in $N/N_\epsilon$, (\ref{1}) is equal to the refined Witten identity displayed in (\ref{refined}). As  a result, (\ref{final2}) may further be elaborated to become
\begin{eqnarray}
&&2\pi(M-|P|)\nonumber\\
&\ge&\,\int_{N_\epsilon}\Big[\frac{1}{4}(\mu u-j^an_a)+\frac{1}{2}u^5\,|\hat D_{AB}\hat\lambda_C|^2\nonumber\\
&&+\frac{1}{4}\Big(\,\frac{1}{8u}\,|\dot
h_{ab}|^2-\frac{1}{2u}|D_{(a}n_{b)}|^2\Big)
 +\frac{1}{4}u(\nu^a-n^a)D_aK.\Big]\nonumber\\
&+&\int_{N/N_\epsilon}\frac{1}{4}(\mu -j^a\nu_a)u+\frac{1}{2}u^5\,|\hat {\mathscr D}_{AB}\hat\lambda_C|^2\nonumber\\
&+&\,\frac{1}{4}\int_{S}\,\,-\sqrt
2\,u^{-3}(\lambda^{\dag
A}\,\cancel\nabla_{A}{}^C\lambda_C\,+\,\lambda_{A}\,\cancel\nabla^{AC}\lambda_C^{\dag})
\label{final20}
\end{eqnarray}
Within $N_\epsilon$, from (\ref{alpha}), we have
\begin{equation}
\label{shift-alp-1} n_a= u(1-\eta)^2\nu_a+\sqrt 2 u\Big[\eta^2\tilde o_{(C}\,\tilde\iota{\hbox{\hskip
-2pt}}_{\,D)}+\eta(1-\eta)(\tilde o_{(C}\,\iota{\hbox{\hskip
-2pt}}_{\,D)}+ o_{(C}\,\tilde\iota{\hbox{\hskip
-2pt}}_{\,D)})\Big].
\end{equation}
By the Cauchy-Schwarz inequality,
\begin{eqnarray}
&&|D_{(a}n_{b)}|^2\nonumber\\
&\le&
 \frac{1}{4}\,\big|\alpha_bD_au+\alpha_aD_bu\big|^2+u^2(1-\eta)^2|D_{(a}\nu_{b)}|^2\nonumber\\
 &&+2u^2\Big |\reta^2\,D_{AB}\tilde o_{(C}\,\tilde\iota{\hbox{\hskip
-2pt}}_{D)}\,+\,\reta\,(1-\reta)\,\big[D_{AB}\tilde o_{(C}\,\iota{\hbox{\hskip
-2pt}}_{\,D)}\,+\,D_{AB}o_{(C}\,\tilde\iota{\hbox{\hskip
-2pt}}_{\,D)}\big]\nonumber\\
 &&
 +\,2\tilde o_{(C}\,\tilde\iota{\hbox{\hskip
-2pt}}_{\,D)}\,\reta
D_{AB}\reta\,-\,2o_{(C}\,\iota\,{\hbox{\hskip
-2pt}}_{\,D)}\,(1-\reta)
D_{AB}\reta\nonumber\\
&&+\,\big[\,\tilde o_{(C}\,\iota{\hbox{\hskip
-2pt}}_{\,D)}+o_{(C}\,\tilde\iota{\hbox{\hskip
-2pt}}_{\,D)}\big]\big[
(1-\reta)D_{AB}\reta-\reta D_{AB}\reta\big]\Big|^2\nonumber \\
&\le&u^2|D_{(a}\nu_{b)}|^2+C_1\nonumber\\
\label{estimate}
\end{eqnarray}
for some constant $C_1$ determined by $\sup_{N_\delta}(\,u,|D_au|,|D_{AB}o_C|, |D_{AB}\tilde o_C|)$ and we have used (\ref{shift-alp-0}) in arriving at the
final inequality.  In a similar way,
\begin{eqnarray}
&&\big|(u\nu^a-n^a)D_aK\big|\nonumber\\
&<&\sqrt 2 u\big|\eta^2\tilde o^A\tilde\iota^B+(2\eta-\eta^2)o^A\iota^B+\eta(1-\eta)(\tilde o^A\iota^B+ o^A\tilde\iota^B)\big|\big|D_{AB}K\big|\nonumber\\
&<&C_2\nonumber\\
\label{shift-alpp}
\end{eqnarray}
for some constant $C_2$ determined by $\sup_{N_\delta}(u, |D_aK|)$.

By construction, $N_\epsilon$ is generated by a one
parameter family of spheres $S_x,x\in [0,\epsilon)$ and denote by $A_x$ the
area of $S_x$, we have from (\ref{estimate}) and (\ref{shift-alpp}) and the foliated structure
of $N_\epsilon$ that, for $\epsilon<\delta$,
\begin{eqnarray}
&&\int_{N_\epsilon}\,\frac{1}{2u}|D_{(a}n_{b)}|^2+(u\nu^a-n^a)D_aK \nonumber\\
&\le&\big(\int_{N_\epsilon}\,\frac{u}{2}|D_{(a}\nu_{b)}|^2\,\big)\,+\,
(C_1+C_2)\int_0^\epsilon  A_x\,dx\nonumber
 \\
 &<&\big(\int_{N_\epsilon}\,\frac{u}{2}|D_{(a}\nu_{b)}|^2\,\big)+(C_1+C_2)\,
\epsilon \,\Big(\sup_{x\in[0,\epsilon]}A_x\,\Big)\nonumber\\
 &<&\big(\int_{N_\epsilon}\,\frac{u}{2}|D_{(a}\nu_{b)}|^2\,\big)+C\,\epsilon\label{epsilon}
\end{eqnarray}
where $C= (C_1+C_2)\,\sup_{x\in[0,\delta]}A_x$.

In view of (\ref{epsilon}), the integral over $N_\epsilon$ in (\ref{final20}) may further be expressed as
\begin{eqnarray}
&&\int_{N_\epsilon}\Big[\frac{1}{4}(\mu u-j^an_a)+\frac{1}{2}u^5\,|\hat D_{AB}\hat\lambda_C|^2+\frac{1}{4}\Big(\,\frac{1}{8u}\,|\dot
h_{ab}|^2-\frac{1}{2u}|D_{(a}n_{b)}|^2\Big)\nonumber\\
&&\phantom{\,\int_{N}\,} +\frac{1}{4}u(\nu^a-n^a)D_aK.\Big]\nonumber\\
&>&\int_{N_\epsilon}\,\Big[\frac{1}{4}(\mu u-j^an_a)+\frac{1}{2}u^5\,|\hat D_{AB}\hat\lambda_C|^2+\frac{1}{4}\Big(\,\frac{1}{8u}\,|\dot
h_{ab}|^2-\frac{u}{2}|D_{(a}\nu_{b)}|^2\Big)-C\epsilon\nonumber\\
&=&\int_{N_\epsilon}\,\Big[\frac{1}{4}(\mu u-j^an_a)+\frac{1}{2}u^5\,|\hat {\mathscr D}_{AB}\hat\lambda_C|^2-C\epsilon
\label{epsilon-0}
\end{eqnarray}
 where the last equality follows from the definition of the Sen-Witten operator together with (\ref{kab0}) with $u\nu_a$ in place of $n_a$ in it. Putting (\ref{epsilon-0}) back into (\ref{final20}), we then find
\begin{eqnarray}
&&2\pi(M-|P|)\nonumber\\
&\ge&\,\int_{N}\frac{1}{4}(\mu u-j^an_a)+\frac{1}{2}u^5\,|\hat {\mathscr D}_{AB}\hat\lambda_C|^2-C\epsilon\nonumber\\
&+&\,\frac{1}{4}\int_{S}\,\,-\sqrt
2\,u^{-3}(\lambda^{\dag
A}\,\cancel\nabla_{A}{}^C\lambda_C\,+\,\lambda_{A}\,\cancel\nabla^{AC}\lambda_C^{\dag}).
\label{final200}
\end{eqnarray}
The term $C\epsilon$ is an additional term to an otherwise manifestly positive volume integal in (\ref{final200})
that
generates by the twisting of $\lambda_A$. This additional term
 may be suppressed to be sufficiently small provided the annular region $N_\epsilon$ is  chosen to be sufficiently small by shrinking $\epsilon$.
  The arbitrariness of $\epsilon$ then means that the positivity of the integral over $N$ is not disturbed.
 With all these considerations,
  we may then infer from (\ref{final200}) that
\begin{eqnarray}
2\pi(M-|P|)
&\ge&\frac{1}{4}\int_{S}\,\,-\sqrt 2\,f^{-4}\,(\lambda^{\dag
A}\,\cancel\nabla_{A}{}^C\lambda_C\,+\,\lambda_{A}\,\cancel\nabla^{AC}\lambda_C^{\dag})\label{positive}
\end{eqnarray}
where for notational convenience later on, we have written
\begin{equation}
\label{f}
f^4=u^3.
\end{equation}
Likewise, in the past trapped case when $N_a$ is chosen to be inward
pointing and given by $N_a=-u\nu^a$, we  deduce in a similiar way the
validity of (\ref{positive}).

\section{Evaluation of the inner boundary term}

In our next step, we shall evaluate the inner boundary term worked out in (\ref{positive}). The presence of $f^{-4}$ in the integrand of (\ref{positive}) means that the calculation will not be entirely straightforward. We will have to appeal to the APS boundary condition
satisfied by $\lambda^A$ in a less obvious way and the arguments are more intricate than originally anticipated.

Consider the following operator
\begin{equation}
\label{operator}
L_{A}{}^{C}=-\cancel\nabla_{A}{}^{C}-\mu_0\epsilon_{A}{}^{C}.
\end{equation}
In order to obtain a lower bound of the inner boundary term in (\ref{positive}) in terms of the areal radius of the marginally trapped surface,
it is sufficient to prove that
\begin{equation}
\label{statement-1}
\int_S dS\,\,f^{-4}\,
(\lambda^{\dagger\,A}L_{A}{}^{C}\lambda_C\,+\,\lambda_{A}L^{AC}\lambda^{\dagger}_C)\,\ge  0.
\end{equation}

To begin with, it is not difficult to see that, when restricted to the Hilbert space spanned by the eignevectors of $\{\mu_n\}$, $L_{A}{}^{C}$ becomes a positive operator and
therefore admits a unique
square root operator $T_{A}{}^{C}$ so that
\begin{equation*}
L_{A}{}^{C}=T_{A}{}^{B}T_{B}{}^{C}.
\end{equation*}
The inner boundary integral in (\ref{statement-1}) may then be further expressed as
\begin{eqnarray}
&&\int_S\,f^{-4}( \lambda^{\dag
A}\,L_{A}{}^C\lambda_C\,+\,\lambda_A\,L^{AC}\lambda^{\dagger}_C)\,\nonumber\\
&=&\int_S dS\,\,f^{-4}\,
(\lambda^{\dagger\,A}T_{A}{}^{M}T_{M}{}^{N}\lambda_N\,+\,\lambda_{A}T^{AM}{}T_{M}{}^{N}\lambda^{\dagger}_N).\label{inner-2}
\end{eqnarray}
The formal analogy between $ L_{A}{}^{C}$ and $ T_{A}{}^{C}$ plus a large amount of  calculations in terms of $ T_{A}{}^{C}$ raise the question
 whether it is feasible to
develop the calculus of $T_{A}{}^{C}$ similiar to that of $L_{A}{}^{C}$.   It turns out that this expectation is not far off the mark
and, perhaps in a way not entirely expected, we need some
holomorphic functional calculus  to realise it.

For a spherical two surface, the inverse operator
$L^{-1}{}_{A}{}^{C}$ exists. It is bounded and  again
positive. It admits a square root operator $T^{-1}{}_{A}{}^{B}$ so
that
\begin{equation*}
L^{-1}{}_{A}{}^{C}=T^{-1}{}_{A}{}^{B}T^{-1}{}_{B}{}^{C}.
\end{equation*}
By the Cauchy integral formula for the analytic function of a
bounded operator,
 $T^{-1}{}_{A}{}^{C}$ admits an integral representation
 \begin{equation}
 \label{inverse-sqrt}
 T^{-1}{}_{A}{}^{C}=\frac{1}{2\pi i}\int_\Gamma dz\,\, z^{-\frac{1}{2}}\,
R_{A}{}^{C},
\end{equation}
 where
 \begin{equation*}
 R_{A}{}^{C}:=\left(z\epsilon_A{}^C\,-\,L_{A}{}^{C}\right)^{-1}
 \end{equation*}
is the resolvent operator of $L_{A}{}^{C}$ defined in
the standard way  and $\Gamma$ is a  contour closed at $\infty$
that encloses the eigenvalues of $L^{-1}{}_{A}{}^{C}$
along the positive real axis. To be concrete, choose the contour
$\Gamma=\cup_{n=0}^{\infty}\,\gamma_n$ so that, for each $n$, $\gamma_n$ is a small circle centered at $\mu_n$
defined by
\begin{equation*}
\gamma_n=\{z\in C|z=-\mu_n+\mu_0+\epsilon\,e^{i\theta}\, \hbox{for some sufficiently small}\, \epsilon\in R \hbox{ and} \,
\theta\in[0,2\pi)\}.
\end{equation*}
 From (\ref{inverse-sqrt}), we
then have (\cite{kato}, Chapter 5, Section 10)
\begin{equation}
\label{sqrt-2}
T_{A}{}^{C}=\frac{1}{2\pi i}\int_\Gamma dz\,\,
z^{-\frac{1}{2}}R_{A}{}^{M}\,L_{M}{}^{C}.
\end{equation}
It then follows from the definition of $R_{A}{}^{C}$ that
it commutes with $L_{A}{}^{C}$ which, in terms of the
index notation, may be written as
\begin{equation}
\label{order}
L_{A}{}^{M}\,R_{M}{}^{C}=R_{A}{}^{M}\,L_{M}{}^{C}.
\end{equation}
Given (\ref{sqrt-2}) and (\ref{order}), we may rewrite (\ref{inner-2}) as
\begin{eqnarray}
&&\int_S\,f^{-4}( \lambda^{\dag
A}\,L_{A}{}^C\lambda_C\,+\,\lambda_A\,L^{AC}\lambda^{\dagger}_C)\nonumber\\
&=&\frac{1}{\,2\pi i}\frac{1}{\,2\pi i}\int_{\Gamma}dz\int_{\Gamma}dw\int_S dS
\quad z^{-\frac{1}{2}}w^{-\frac{1}{2}}\,f^{-4}\,\,\big[\lambda^{\dagger\,A}R_{A}{}^{M}L_{M}{}^{B}L_{B}{}^{C}R_{C}{}^{D}\lambda_{D}\nonumber\\
&&\phantom{\frac{1}{\,2\pi i}\frac{1}{\,2\pi i}\int_{\Gamma}dz\int_{\Gamma}dw\int_S dS}+\,\lambda_{A}R^{AM}L_{M}{}^{B}L_{B}{}^{C}R_{C}{}^{D}\lambda^{\dagger}_D\big].\label{inner-3}
\end{eqnarray}
This suggests to us to define
\begin{equation}
\label{def-omega}
\omega_A=\frac{1}{2\pi i}\int_\Gamma dz\,\,
z^{-\frac{1}{2}}R_{A}{}^{M}\,\lambda_{M}.
\end{equation}
In terms of spectral representation of the resolvent operator $R_{A}{}^{C}$ given as
\begin{equation*}
R_{A}{}^{C}=\sum_{n=0}^{\infty}\frac{1}{ z-\mu_n}\lambda_{n}{}_{A}\lambda_{n}{}^{\dag C},
\end{equation*}
it may  be checked that
\begin{equation*}
\omega^\dagger_A=\frac{1}{2\pi i}\int_\Gamma dz\,\,
z^{-\frac{1}{2}}R_{A}{}^{M}\,\lambda^\dagger_{M}.
\end{equation*}
(\ref{inner-3}) may then be written in a more compact form as
\begin{eqnarray}
&&\int_S\,f^{-4}( \lambda^{\dag
A}\,L_{A}{}^C\lambda_C\,+\,\lambda_A\,L^{AC}\lambda^{\dagger}_C)\,\nonumber\\
&=&\int_S dS\,\,f^{-4}\,
(\omega^{\dagger\,A}L_{A}{}^{M}L_{M}{}^{N}\omega_N\,+\,\omega_{A}L^{AM}{}L_{M}{}^{N}\omega^{\dagger}_N)\label{integrand}
\end{eqnarray}
From this point on, we may evaluate the integrand in (\ref{integrand}) in terms of standard spinor calculus. From the definition of $L_{A}{}^{M}$ in (\ref{operator}), we have
\begin{eqnarray}
&&f^{-4}\,\omega^{\dagger\,A}L_{A}{}^{M}L_{M}{}^{N}\omega_N\nonumber\\
&=&f^{-4}\,\omega^{\dagger\,A}(-\cancel\nabla_{A}{}^{M}-\mu_0\epsilon_{A}{}^{M})
(-\cancel\nabla_{M}{}^{N}-\mu_0\epsilon_{M}{}^{N})\omega_N\nonumber\\
&=&f^{-4}\,\big[\omega^{\dagger\,A}\cancel\nabla_{A}{}^{M}\cancel\nabla_{M}{}^{N}\omega_N\,
+\,2\mu_0\,\omega^{\dagger\,A}\cancel\nabla_{A}{}^{M}\omega_M\,+\,\mu_0^2\,\omega^{\dagger\,A}\omega_A\big]\nonumber\\
&=&f^{-4}\big[\,\cancel\nabla_{A}{}^{M}(\omega^{\dagger\,A}\cancel\nabla_{M}{}^{N}\omega_N)\,-\,
({\cancel{\nabla}}_C{}^N\omega^{\dag\,C})\,(\cancel{\nabla}_N{}^M\,\omega_{M})\,\nonumber
\\&&\phantom{f^{-4}\,}+\,2\mu_0\,\omega^{\dagger\,A}
\cancel\nabla_{A}{}^{M}\omega_M
+\mu_0^2\,\omega_A\omega^{\dagger\,A}\big].
\label{term by term}
\end{eqnarray}
Likewise, the term $f^{-4}\,\omega_AL^{AM}L_{M}{}^{N}\omega^{\dagger}_N$ in (\ref{integrand}) may be calculated in a similar manner and together with (\ref{term by term}) we have
\begin{eqnarray}
&&f^{-4}\,(\omega^{\dagger\,A}L_{A}{}^{M}L_{M}{}^{N}\omega_N\,+\,\omega_A L^{AM}L_{M}{}^{N}\omega^{\dagger}_N)
\nonumber\\
&=&f^{-4}\,\big[\cancel\nabla_{A}{}^{M}(\omega^{\dagger\,A}\cancel\nabla_{M}{}^{N}\omega_N)
+\cancel\nabla^{AM}(\omega_A\cancel\nabla_{M}{}^{N}\omega^{\dagger}{}_N)\nonumber\\
&&-2({\cancel{\nabla}}_C{}^N\omega^{\dag\,C})\,(\cancel{\nabla}_N{}^M\,\omega_{M})\nonumber\\
&&+\,2\mu_0\,(\omega^{\dagger\,A}\cancel\nabla_{A}{}^{M}\omega_M\,+\,\omega_A\cancel\nabla^{AM}
\omega^{\dagger}{}_M)
+\,2\mu_0^2\,\omega_A\omega^{\dagger\,A}\big].
\label{abd}
\end{eqnarray}
We shall now evaluate (\ref{abd}) term by term.
Define
\begin{equation*}
\label{definition}
  \beta^A=f^{-2}\omega^A, \quad
 \beta^{\dag\,A}=f^{-2}\omega^{\dag\,A}.
 \end{equation*}
 Consider first
\begin{eqnarray}
&&f^{-4}\,{\cancel{\nabla}}_C^{N}(\omega^{\dag C}\,\cancel{\nabla}_N{}^M\,\omega_{M})\nonumber\\
&=&f^{-4}\,{\cancel{\nabla}}_C{}^N(f^{2}\beta^{\dag\,C}\,\cancel{\nabla}_N{}^M\,f^{2}\beta_{M})\nonumber\\
&=&f^{-4}\,{\cancel{\nabla}}_C{}^N[f^{2}\beta^{\dag\,C}\,(\beta_M\cancel{\nabla}_N{}^M\,f^{2}+f^{2}\cancel{\nabla}_N{}^M\beta_{M})]\nonumber\\
&=&-\frac{1}{2}f^{-4}\,{\cancel{\nabla}}_C{}^N(\beta^{\dag\,C}\beta^M\cancel{\nabla}_{MN}\,f^{4})+
f^{-4}\,{\cancel{\nabla}}_C{}^N(f^{4}\beta^{\dagger C}\cancel{\nabla}_N{}^M\beta_{M})\nonumber\\
&=&-\frac{1}{2}f^{-4}\,{\cancel{\nabla}}_C{}^N(\beta^{\dag\,C}\beta^M\cancel{\nabla}_{MN}\,f^{4})\,+{\cancel{\nabla}}_C{}^N(\beta^{\dagger C}\cancel{\nabla}_{NM}\beta^{M})\nonumber\\
&&+4\,(\beta^{\dagger C}{\cancel{\nabla}}_C{}^N\ln f)(\cancel{\nabla}_N{}^M\beta_{M}).
\label{first}
\end{eqnarray}
 Likewise, the term $f^{-4}\,{\cancel{\nabla}}^{AM}(\omega_A\,\cancel{\nabla}_M{}^N\omega^{\dag}{}_ N) $ may be calculated in a similar way and we find
\begin{eqnarray}
&&f^{-4}\,{\cancel{\nabla}}^{AN}(\omega_A\,\cancel{\nabla}_M{}^N\omega^{\dag}{}_ N)\nonumber\\
&=&-\frac{1}{2}f^{-4}\,{\cancel{\nabla}}_C{}^N(\beta^{\dag\,C}\beta^M\cancel{\nabla}_{MN}\,f^{4})\,+{\cancel{\nabla}}_C{}^N(\beta^{\dagger C}\cancel{\nabla}_{NM}\beta^{M})\nonumber\\
&&+4\,(\beta^{\dagger C}{\cancel{\nabla}}_C{}^N\ln f)(\cancel{\nabla}_N{}^M\beta_{M}).\label{second}
\end{eqnarray}
Adding up (\ref{first}) and (\ref{second}), we have
\begin{eqnarray}
&&f^{-4}\,\big[\cancel\nabla_{A}{}^{M}(\omega^{\dagger\,A}\cancel\nabla_{M}{}^{N}\omega_N)
+\cancel\nabla^{AM}(\omega_A\cancel\nabla_{M}{}^{N}\omega^{\dagger}{}_N)]\nonumber\\
&=&\,{\cancel{\nabla}}_C{}^N(\beta^{\dagger C}\cancel{\nabla}_N{}^M\beta_{M})+\,{\cancel{\nabla}}^{AM}(\beta_A\cancel{\nabla}_M{}^N\beta^{\dag}{}_{N})
\nonumber\\
&&+4(\,\beta^{\dagger C}{\cancel{\nabla}}_C{}^N\ln f)(\cancel{\nabla}_N{}^M\beta_{M})+4(\,\beta_A{\cancel{\nabla}}^{AM}\ln f)(\cancel{\nabla}_M{}^N\beta^{\dag}{}_{N}).
\label{first two terms}
\end{eqnarray}
For the third term on the right hand side of (\ref{abd}),
\begin{eqnarray}
&&2f^{-4}\,({\cancel{\nabla}}_C{}^N\omega^{\dag\,C})\,(\cancel{\nabla}_N{}^M\,\omega_{M})\nonumber\\
&=&2f^{-4}\,({\cancel{\nabla}}_C{}^Nf^{2}\beta^{\dag\,C})\,(\cancel{\nabla}_N{}^M\,f^{2}\beta_{M})\nonumber\\
&=&2f^{-4}\,[\beta^{\dag\,C}{\cancel{\nabla}}_C{}^Nf^{2}+f^{2}{\cancel{\nabla}}_C{}^N\beta^{\dag\,C})]
\,[\beta_{M}\cancel{\nabla}_N{}^M\,f^{2}+f^{2}\cancel{\nabla}_N{}^M\,\beta_{M}]\nonumber\\
&=&2({\cancel{\nabla}}_C{}^N\beta^{\dag\,C})(\,\cancel{\nabla}_N{}^M\,\beta_{M})+8(\beta^{\dag\,C}{\cancel{\nabla}}_C{}^N\ln
f)(\beta_{M}\cancel{\nabla}_N{}^M\,\ln f)\nonumber\\
&&+4(\beta_{M}\cancel{\nabla}_N{}^M\, \ln
f\,)({\cancel{\nabla}}_C{}^N\beta^{\dag\,C})+4(\beta^{\dag\,C}{\cancel{\nabla}}_C{}^N\ln
f)(\cancel{\nabla}_N{}^M\,\beta_{M}).\label{third}
\end{eqnarray}
Next consider the term
\begin{eqnarray}
&&2\mu_0f^{-4}\,(\omega^{\dagger\,A}\cancel\nabla_A{}^{M}\omega_M\,+\,\omega_{A}\cancel\nabla^{AM}\omega^{\dagger}{}_M)\nonumber\\
&=&2\mu_0f^{-4}\,(f^2\beta^{\dagger\,A}\cancel\nabla_{A}{}^{M}f^2\beta_M\,+\,f^2\beta_{A}\cancel\nabla^{AM}f^2\beta^{\dagger}{}_M)
\nonumber\\
&=&2\mu_0(\beta^{\dagger\,A}\cancel\nabla_{A}{}^{M}\beta_M+\beta_{A}\cancel\nabla^{AM}\beta^{\dagger}{}_M\nonumber\\
&&+
2\beta_M\beta^{\dagger A}\cancel\nabla_{A}{}^{M}\ln f+2\beta_M\beta^{\dagger}{}_A\cancel\nabla^{AM}\ln f)\nonumber\\
&=&2\mu_0(\beta^{\dagger\,A}\cancel\nabla_{A}{}^{M}\beta_M+\beta_{A}\cancel\nabla^{AM}\beta^{\dagger}{}_M\nonumber\\
&&-
2\beta^M\beta^{\dagger A}\cancel\nabla_{AM}\ln f+2\beta^M\beta^{\dagger A}\cancel\nabla_{AM}\ln f)\nonumber\\
&=&2\mu_0(\beta^{\dagger\,A}\cancel\nabla_{A}{}^{M}\beta_M+\beta_{A}\cancel\nabla^{AM}\beta^{\dagger}{}_M).
\label{dirac00}
\end{eqnarray}
Substituting (\ref{first two terms}), (\ref{third}) and (\ref{dirac00}) back into (\ref{abd}), we see that many terms not manifestly positive in (\ref{first two terms}) and (\ref{third}) mutually cancel each other and we finally have
\begin{eqnarray}
&&f^{-4}\,
(\omega^{\dagger\,A}L_{A}{}^{M}L_{M}{}^{N}\omega_N\,+\,\omega_{A}L^{AM}{}L_{M}{}^{N}\omega^{\dagger}_N)\nonumber\\
&=&| \cancel\nabla_{A}{}^{M}\beta_M|^2   + 2\mu_0(\beta^{\dagger\,A}\cancel\nabla_{A}{}^{M}\beta_M+\beta_{A}\cancel\nabla^{AM}\beta^{\dagger}{}_M)
+2\mu_0|\beta|^2\nonumber\\        &&+{\cancel{\nabla}}_C{}^N(\beta^{\dagger C}\cancel{\nabla}_N{}^M\beta_{M})+\,{\cancel{\nabla}}^{AM}(\beta_A\cancel{\nabla}_M{}^N\beta^{\dag}{}_{N})\nonumber\\
&=&[(-\cancel\nabla_{A}{}^{M}-\mu_0\epsilon_{A}{}^{M})\beta_M]^\dag\,
[(-\cancel\nabla_{M}{}^{N}-\mu_0\epsilon_{M}{}^{N})\beta_N]
\nonumber\\
&&+{\cancel{\nabla}}_C{}^N(\beta^{\dagger C}\cancel{\nabla}_N{}^M\beta_{M})+\,{\cancel{\nabla}}^{AM}(\beta_A\cancel{\nabla}_M{}^N\beta^{\dag}{}_{N})\nonumber\\
&=&|L_M{}^N\omega_N|^2+{\cancel{\nabla}}_C{}^N(\beta^{\dagger C}\cancel{\nabla}_N{}^M\beta_{M})+\nonumber\\&&\,{\cancel{\nabla}}^{AM}(\beta_A\cancel{\nabla}_M{}^N\beta^{\dag}{}_{N})
\label{div}
\end{eqnarray}
according to the definition of $L_M{}^N$ given in (\ref{operator}).
When integrating (\ref{div}) over $S$, the divergence terms in (\ref{div}) vanish  and we get from (\ref{integrand}) and (\ref{div})  that
\begin{eqnarray}
&&\int_S\,f^{-4}( \lambda^{\dag
A}\,L_{A}{}^C\lambda_C\,+\,\lambda_A\,L^{AC}\lambda^{\dagger}_C)\nonumber\\
&&\int_S\,f^{-4}\,
(\omega^{\dagger\,A}L_{A}{}^{M}L_{M}{}^{N}\omega_N\,+\,\omega_{A}L^{AM}{}L_{M}{}^{N}\omega^{\dagger}_N)\nonumber\\
&=& \int_S\,
|L_M{}^N\omega_N|^2\,>\,0\nonumber\\\label{operator-1}
\end{eqnarray}
as desired. From (\ref{operator}) and (\ref{operator-1}),  we may infer
\begin{eqnarray*}
&&-\int_S\,f^{-4}( \lambda^{\dag
A}\,\cancel\nabla_{A}{}^C\lambda_C\,+\,\lambda_A\,\cancel\nabla^{AC}\lambda^{\dagger}_C)\nonumber\\
&\ge& 2\mu_0\int_S\,f^{-4}\,\lambda_A \lambda^{\dag A}\nonumber\\
&=&2\mu_0\int_S\,u
\end{eqnarray*}
according to the definition stated in (\ref{fourth}) and (\ref{f}). (\ref{positive}) then becomes
\begin{eqnarray}
M-|P|&\ge& c\,r\label{inequality-final}
\end{eqnarray}
where we have used $\mu_0\ge\frac{1}{r}$ for a spherical surface [3] and
\begin{equation}
\label{definition-c}
  c=\inf_S \,u.
 \end{equation}

So far we have been assuming that $u$ is strictly positive. This hypothesis may be relaxed by a suitable regulariation (or cutoff) of the zero points of $\lambda_A$ and details will be presented in Section 9. As the final step, we shall estimate the upper bound of the  constant $c$ to complete the proof.

Suppose for some
$x\in\partial N$, $\lambda_A=0$. The APS boundary condition in
(\ref{aps1}) then implies that $\lambda_A$ vanishes everywhere  in $\partial N$ and $\partial N$ is a set of zero points of infinite order.
 Subject to
(\ref{sw}), we have the following elliptic system
\begin{equation}
\label{elliptic}
{\mathscr{D}}^2\lambda_A=\frac{1}{2}(\mu\,\epsilon_A{}^L-j_A{}^L)\lambda_L, \quad
A=0,1.
\end{equation}
 where $\mathscr{D}^2=-\mathscr{D}_{AB}{\mathscr{D}}^{AB} $ is a generalised Laplacian.
It may be checked that $|\mathscr{D}^2\lambda_A|\le C |\lambda_A|$ for some constant $C$.
For a sufficiently small coordinate ball $B$ centered
at $x$, standard reflection across $\partial N\cap B$ enables us to
extend (\ref{elliptic}) from $B\,\cap\, R_+^3$ to the entire $B$ as
an elliptic system with Lipshitz coefficients. Unique
continuation at the point $x$ then implies $\lambda_A=0$ everywhere in  $N$ (see \cite{kazdan}, Theorem 1.8) and
this contradicts the
asymptotic boundary condition satisfied by $\lambda_A$ near
infinity. We may then infer $c>0$ and the inequality in (\ref{inequality-final}) is not vacuous.

To estimate the upper bound of  $c$, we revert to the spinor norm $\rvarphi=u^4$ and we have
\begin{eqnarray}
\mathscr{D}^2\rvarphi&=&D^a\left(D_a\varphi\,-\,K_{ab}\nu^a\varphi\right)\nonumber\\
&=&(\mu-j_a\nu^a)\,\varphi\,+\,2\,|\mathscr{D}_{AB}\lambda_C|^2\label{divergence}
\end{eqnarray}
where $\nu^a=\sqrt 2\,o_{(A}\iota_{B)}$.  The dominant energy condition implies that
 $\mathscr{D}^2\rvarphi>0$. Further, (\ref{divergence}) is an elliptic PDE of divergence form. The maximum principle \cite{trud} applies and the maximum of $\rvarphi$ will occur either at inner boundary $S$ or at infinity.
Suppose on the contrary that the maximum of $\rvarphi$ occurs at some $x\in S$. It  follows that $\frac{\partial\varphi}{\partial \nu}< 0$ at $x$. Continuity implies there exists a neighhourhood $U\subset S$  centered at $x$ such that $\frac{\partial\varphi}{\partial \nu}< 0$ in $U$.   Fix a  cutoff function $\eta>0$ in $U$, by twisting $\lambda_A$ in an appropriate way as before, the integral form of (\ref{divergence}) together with the Sen-Witten equation in (\ref{sw}) and the marginally trapped condition on $S$ give
\begin{eqnarray}
\int_S\,\eta^4\frac{\partial\rvarphi}{\partial\nu}
&=&-\int_S\,\sqrt 2\,\eta^4(\lambda^{\dag
A}\,\cancel\nabla_{A}{}^C\lambda_C\,+\,\lambda_{A}\,\cancel\nabla^{AC}\lambda_C^{\dag})
\label{integral2}
\end{eqnarray}

 Given the APS boundary condition, $-\cancel\nabla_M{}^N$ is a positive operator and admits a unique square root operator. With $\eta^4$ and $-\cancel\nabla_M{}^N$ in place of $u^{-3}$ and $L_M{}^N$ respectively in (\ref{integrand}) and by repeating the arguments leading to (\ref{operator-1}), we have $\int_S\,\eta^4\frac{\partial\varphi}{\partial\nu}>0$
 and this contradicts our initial hypothesis that
 $\frac{\partial\varphi}{\partial \nu}< 0$ at $U$.  Therefore, the maximum of $ \rvarphi $ will  occur at the asymptotic regime and
we  necessarily have $c<1$. As a result, we have $1>c>0$ in (\ref{inequality-final}).

 With the trivial generalisation to the case of  multi-connected horizon, we are then finally in a position to state the following theorem.

\vskip 10pt
\par\noindent {\bf Theorem. }
\vskip  10pt

 {\it Let $(N, h_{ab}, K_{ab})$ be an asymptotically Euclidean initial
 data set with inner boundary $\partial N=\cup_{i=0}^{n-1}\,S_i$,
 where $S_i,\,i=0,.. n-1$ are disjoint, smooth future or past marginally trapped surfaces with spherical topology
 and areal radius $r_i$. Subject to the dominant
energy condition, we  have
\begin{equation*}
M-|P|\ge\,c\,\sum_{i=1}^n\,r_i,\quad 0<c<1.
\end{equation*}

}

\section{Derivation of the refined Witten identity}

We will now go back to fill in certain details in the steps leading to the proof of the theorem just stated.
In this section, we shall first provide more details on the derivation of the
spinor identity stated in (\ref{finn}).

The crux of the calculations leading to (\ref{finn})  is  to evaluate the term
\begin{equation}
\label{eq} |D_{AB}o_C|^2=-D_{AB}o_C \, D^{AB}\iota^C.
\end{equation}
Given (\ref{11}), we have
\begin{eqnarray}
\label{aaa}
&&D_{AB}o_C\,D^{AB}\iota^C\nonumber\\
 &=& D_{AB}(u^{-1}\hat \lambda_C)
D^{AB}(u\hat\lambda^{\dagger\,C})\nonumber\\ &=& \left(-\hat
\lambda_C\,u^{-2}D_{AB}u +u^{-1}D_{AB} \hat \lambda_C \right)\nonumber\\
&&\left(\hat\lambda^{\dagger\,C}
D^{AB}u+uD^{AB}\hat\lambda^{\dagger\,C}\right).
\end{eqnarray}
It is standard to work out that, under the conformal rescaling
$h_{ab}\rightarrow\hat h_{ab}=u^4h_{ab}$,
\begin{equation} \label{conf1}
\begin{split}
\hat D_{AB}\, \hat \lambda_C&= D_{AB} \hat \lambda_C-\hat \lambda_BD_{CA}\ln u-
 \hat \lambda_AD_{CB}\ln
 u\\
 \hat D_{AB} \,\hat \lambda^{\dagger\,\,C}&= D_{AB}\, \hat \lambda^{\dagger\,\,C}
 +\epsilon_A{}^C\,\hat \lambda^{\dagger\,\,M}D_{BM}\ln u
 +\epsilon_B{}^C\,\hat \lambda^{\dagger\,\,M}D_{AM}\ln u
\end{split}
\end{equation}
Substitute (\ref{conf1})  into (\ref{aaa}), we then
have
\begin{eqnarray}
&&D_{AB}o_C\,D^{AB}\iota^C\nonumber\\
 &=&\left(u^{-1}\hat D_{AB}
\hat\lambda_C+u^{-2}\hat
\lambda_BD_{CA} u+u^{-2}\hat\lambda_AD_{CB} u-u^{-2}\hat \lambda_CD_{AB}u\right)\nonumber\\
&&\left(u\hat D^{AB}\, \hat \lambda^{\dagger\,C}-\epsilon^{AC}\,\hat
\lambda^{\dagger\,M}D^B{}_{M} u-\epsilon^{BC}\,\hat
\lambda^{\dagger\,M}D^A{}_{M}
u+\hat\lambda^{\dagger\,C}D^{AB}u\right)\nonumber\cr
&=&\phantom{+}\hat D_{AB} \hat \lambda_C\,\hat  D^{AB}\hat
\lambda^{\dagger\,C}\nonumber\cr &&+u^{-1}\hat D_{AB} \hat
\lambda_C\left(-\epsilon^{AC}\,\hat \lambda^{\dagger\,M}D^B{}_{M}
u-\epsilon^{BC}\,\hat \lambda^{\dagger\,M}D^A{}_{M}
u+\hat\lambda^{\dagger\,C}D^{AB}u\right)\nonumber\cr &&+ u\hat
D^{AB} \,\hat \lambda^{\dagger\,C}\left(u^{-2}\hat \lambda_BD_{CA}
u+u^{-2}\hat \lambda_AD_{CB} u-u^{-2}\hat
\lambda_CD_{AB}u\right)\nonumber\cr &&+\left(u^{-2}\hat
\lambda_BD_{CA} u+u^{-2}\hat \lambda_AD_{CB} u-u^{-2}\hat\lambda_CD_{AB}u\right)\nonumber\\
&&\phantom{+}\left(-\epsilon^{AC}\,\hat
\lambda^{\dagger\,M}D^B{}_{M} u-\epsilon^{BC}\,\hat
\lambda^{\dagger\,M}D^A{}_{M}
u+\hat\lambda^{\dagger\,C}D^{AB}u\right).\label{comp}
\end{eqnarray}
Subject to the  Sen-Witten equation together with its conformal symmetries, after some standard calculations, we have
\begin{eqnarray}
&&u^{-1}\hat D_{AB} \hat
\lambda_C\left(-\epsilon^{AC}\,\hat \lambda^{\dagger\,M}D^B{}_{M}
u-\epsilon^{BC}\,\hat \lambda^{\dagger\,M}D^A{}_{M}
u+\hat\lambda^{\dagger\,C}D^{AB}u\right)\nonumber\\
&=&(D^{AB}\ln u)(\,\hat \lambda^{\dagger\,C}\hat
D_{AB}\hat \lambda_C\,+\,\frac{1}{ 2}\,
K\,\nu^aD_{a}\ln u)\label{term11}
\end{eqnarray}
Further, using the identity
\begin{eqnarray*}
{  D}_{AB} \lambda^{\dagger}_{C}&=&{  D}_{A(B}\lambda^{\dagger}{}_{C)}
\,+\,{ D}_{A[B} \lambda^{\dagger}{}_{C]}\,,\nonumber\\
&=&{  D}_{A(B}\lambda^{\dagger}{}_{C)}
\,+\,\frac{1}{2}\epsilon_{BC}{ D}_{AN} \lambda^{\dagger N}
\end{eqnarray*}
together with again the Sen-Witten equation  and its conformal symmetries, we may work out
\begin{eqnarray}
&&u\hat D^{AB} \hat
\lambda^{\dagger\,\,C}\left(u^{-2}\hat \lambda_BD_{CA} u+u^{-2}\hat
\lambda_AD_{CB} u-u^{-2}\hat \lambda_CD_{AB}u\right)\nonumber\\
&=&(D_{AB}\ln u)\,(\hat  \lambda_C\hat D^{AB}\hat
\lambda^{\dagger\,C})\,+\,\frac{1}{ 2} K\nu^aD_{a}
\ln u\nonumber\\
&&\label{term22}
\end{eqnarray}
Since $\hat\epsilon^{AB}\hat  \lambda_A\hat \lambda^{\dagger}{}_B=1$, summing terms in
(\ref{term11}) and (\ref{term22}), we have
\begin{eqnarray}
 &&u^{-1}\hat D_{AB} \hat
\lambda_C\left(-\epsilon^{AC}\,\hat \lambda^{\dagger\,M}D^B{}_{M}
u-\epsilon^{BC}\,\hat \lambda^{\dagger\,M}D^A{}_{M}
u+\hat\lambda^{\dagger\,C}D^{AB}u\right)\nonumber\cr
&+&
 u\hat
D^{AB} \,\hat \lambda^{\dagger\,C}\left(u^{-2}\hat \lambda_BD_{CA}
u+u^{-2}\hat \lambda_AD_{CB} u-u^{-2}\hat
\lambda_CD_{AB}u\right)\nonumber\\
&=& K\,\nu^aD_{a} \ln u\nonumber\\
\label{comp1}
\end{eqnarray}
To evaluate in (\ref{comp}) the term
\begin{equation} \label{term3}
\begin{split}
\left(u^{-2}\hat
\lambda_BD_{CA} u+u^{-2}\hat \lambda_AD_{CB} u-u^{-2}\hat \lambda_CD_{AB}u\right)\hbox{\phantom{abcdef}}\\
\hbox{\phantom{abcd}}\left(-\epsilon^{AC}\,\hat \lambda^{\dagger\,\,M}D^B{}_{M}
u-\epsilon^{BC}\,\hat \lambda^{\dagger\,\,M}D^A{}_{M}
u+\hat\lambda^{\dagger\,\,C}D^{AB}u\right),
\end{split}
\end{equation}
 further calculations enable us to infer that
(\ref{term3}) is equal to
\begin{equation}
\label{term33} 6u^{-2}(o^BD_{AB}u)(\iota^ND^A{}_{N}
u)+u^2D_auD^au\,.
\end{equation}
From the Newman-Penrose tetrad constructed from  the spin frame
$(o^A,\iota^A)$,
 a moving three frame  intrinsic
to $N$ may be defined as
\begin{equation*}
m^a =o^A\iota^{A'},\quad \bar{m}^a=o ^{A'}\iota ^A,\quad \nu^a=\frac
{1}{\sqrt{2}}( o^Ao^{A'}-\iota^A\iota^{A'})\,. \label{frame}
\end{equation*}
In terms of $(\nu^a, m^a, \bar{m}^a)$, we have
\begin{equation}
 o^BD_{AB}u
=-\frac{1}{\sqrt 2}\,(\nu^aD_a u)\, o_A-(m^aD_a u)\,\iota_A
\label{term111}
\end{equation}
and
\begin{equation}
\iota^ND^A{}_{N} u=-(\bar m^aD_{a} u)\,o^A+\frac{1}{\sqrt
2}\,(\nu^aD_a u) \,\iota^A\,.\label{term222}
\end{equation}
Using $h_{ab}=\nu_a\nu_b\,+\,2m_{(a}\bar m_{b)}$, we may deduce from
(\ref{term111}) and (\ref{term222}) that
\begin{equation}
\label{term44} u^{-2}(o^BD_{AB}u)(\iota^ND^A{}_{N}
u)=-\frac{1}{2}u^{-2}D_auD^au. \end{equation}
 Therefore we finally
obtain from (\ref{term33}) and (\ref{term44}) that the term in
(\ref{term3}) is equal to $2u^{-2}D_auD^au$. Putting all these together with (\ref{comp1}) back
 to (\ref{comp}), we then have
 \begin{eqnarray}
 &&|D_{AB}o_C|^2\nonumber\\
 &=&- {{D}}_{AB}o_C\,{{D}}^{AB}\iota^C\nonumber\\
 &=&
 u^4\,|\hat {{D}}_{AB} \hat \lambda_C|^2+2|D_a\ln u|^2-K\nu^aD_a\ln u\,\label{fin}
\end{eqnarray}
which is the spinor identity stated in (\ref{finn}). Note that $|\hat {{D}}_{AB} \hat \lambda_C|^2$ is evaluated in terms of
the conformally rescaled symplectic form $\hat\epsilon_{AB}$.

\section{Regularisation of  zero points of a spinor field}

We shall now outline a way to relax the provisional
hypothesis that $\lambda^A$ is non-zero
everywhere in $N$. Given the APS boundary condition, zero points of $\lambda_A$ stay away from the inner boundary  $S$. Denote by $X\subset N/\partial N$ the set of zero points of finite order. The asymptotic boundary condition for $\lambda^A$
means
that $X$ is a subset of some compact set in $N$.
 $X$ is closed then further implies that $X$ is  compact.

It is also known that $X$ is  contained in a countable union of
smooth curves in $N$ \cite{bar1}. Compactness of $X$ implies that
$X\subset \bigcup_{\,k=1}^{\,n} C_k$ for some
natural number $n$ and
$C_k:[0,1]\rightarrow N$ for $k=1,..n$ are smooth curves.  A smooth tubular neighbourhood $T_k:[0,L_k]\times
D_\epsilon \rightarrow N $ may be constructed so that $C_k\subset T_k$, $D_\epsilon$
is a geodesic disk of radius $\epsilon$ centered at a point in
$C_k$.
 In place of $N$, we consider instead
\begin{equation*}N'=N/\ \,\{\hbox{interior
 of}\,\cup\, T_k\}.\end{equation*}
 The integral in (\ref{final2}) then acquires extra boundary terms
 $$\sum_0^k\int_{\partial T_k}\left(\frac{\partial u}{\partial \nu}-uK_{ab}r^a\nu^b\right)$$ where
 $r^a$ is the normal to  $\partial T_k$.

For a zero point $x\in N/\partial N$,   both $u$ and $D_au$
vanish at $x$ and therefore
\begin{equation*}u=O(r^{1/2}),\quad \frac{\partial u}{\partial r} = O(r^{-1/2})\end{equation*}
in $D_\epsilon$ where $r$ is the geodesic distance from $C_k$. Using the compactness of
$\cup_{k=1}^{\,n}\,T_k$ and by means of further calculations, we have
 \begin{eqnarray}
  \left |\,\int_{ \partial T_k} \frac{\partial  u}{\partial r}\,- \,\frac{1}{4}
u\,K_{ab}r^a\nu^b\right |
\leq \alpha  \epsilon^{1/2}.\label{ll}
 \end{eqnarray}
 for some constant $\alpha$ independent of $\epsilon$.
In view of (\ref{ll}), the integral form of the refined Witten identity then becomes
\begin{eqnarray}
&&2\pi(M-|P|)\nonumber\\
&=&\int_{N'} \Delta u-\,\frac{1}{4}
D^a(K_{ab}N^b)   \nonumber\\
&+&\int_{S} \left(\frac{\partial u}{\partial \nu} -K_{ab}N^a\nu^b\right)+o(\epsilon^{1/2}).\label{int2}
 \end{eqnarray}
By shrinking the radius of the tubes $T_k,k=1,\cdots n $ to a
sufficiently
 small $\epsilon$, we see that  the standard positivity argument continues to hold for (\ref{int2})
 when   zero points of $\lambda_A$ are taken into consideration.

\section{Existence and uniqueness of Sen-Witten spinor field and the APS boundary condition.}

We will complete the proof of the above stated theorem by proving
the existence and uniqueness of solution to the Sen-Witten equation
in (\ref{sw}), subject to the APS boundary condition and the
asymptotic boundary condition displayed in (\ref{asy-boundary}).
Once we realise that a suitable amount of twisting of a spinor field described in
(\ref{alpha}) will not disturb the positivity argument,
 the proof becomes quite standard elliptic estimates in terms of the Lax-Milgram approach.  For completeness, we shall briefly sketch it here.

 Denote by $N_R$ the subset in $N$ bounded by a coordinate ball $B_R$ of Euclidean radius $R$ near
 infinity. Fix a real valued function  $\sigma$  in $N$ such that
 $\sigma\ge 1$ and $\sigma=1$ in $N$, $\sigma=r$ in $N/N_{2R}$ where $r$ is the Euclidean radial distance in the
 asymptotic regime. Let  $W_\delta^{k,p}$  be the weighted Sobolev
 spaces defined in the standard way\cite{LP} with $p=2$ and we define the norm of $W_\delta^{k,p}$ in terms of ${{\mathscr D}}_{AB}$.
 Denote the weighted Sobolev norm of $W_{-1}^{1,2}$ by $||\,||$. It is also sufficient to define $||\,||$ in terms of ${{\mathscr D}}_{AB}$ alone \cite{parker}.
  Further restrict the domain of the Sen-Witten operator ${{\mathscr D}}_A{}^C$ to a closed space
  $H_-\subset W_{-1}^{1,2}$ such that $\psi_A\in H_{-}$ if and only if $\psi_A\in  W_{-1}^{1,2} $ and $\psi_A|_{S}$ satisfies
the APS boundary condition given in  (\ref{aps2}) and (\ref{aps1}).

Extend the covariantly constant spinor $\lambda_{\,0A}$ near infinity
in an obvious way to $N$ and denote it by $\eta_A$.  Fix a sequence
of Euclidean radius $R_i$ near
 infinity indexed by natural numbers with $R_{i+1}>R_i$ for all $i$ and
 $\lim_{i\rightarrow\infty}R_i\rightarrow \infty$. Then consider  a
 sequence $\{\eta_{iA}\}$ with support in $N/R_i$ such that
 $\lim_{i\rightarrow\infty}\eta_{iA}\rightarrow \eta_A$.
Define
\begin{equation}
\label{deff}\lambda_{iA}=\psi_{iA}+\eta_{iA}.
\end{equation}
For notation convenience, the $i$ th dependence of $\psi_A$ and
$\eta_A$ will be suppressed in what follows and $\psi_A$ is  assumed
to have support in $N_{R_i}$.

Subject to the dominant energy condition,  $\mathscr D_A{}^C$ is injective. It is then sufficient to consider the
following elliptic operator
\begin{equation}
\label{second-order}
\mathscr{D}_A{}^C\mathscr{D}_C{}^N\psi_N=-\mathscr{D}_A{}^C\mathscr{D}_C{}^N\eta_N
\end{equation}
with the prescribed APS boundary condition at the inner boundary and the asymptotic fall off near spatial infinity.

As in the standard Lax-Milgram approach, define a bilinear form in $H_-$ as
\begin{equation}
\label{bilinear} a( \alpha, \lambda)=\int_N
\,({{\mathscr{D}}}^{CN}\alpha_{ N})^\dag({{\mathscr{D}}}_{C}{}^L\lambda_L)
\end{equation}
 together with the
linear functional in $H_{-}$ defined by
\begin{equation*}
f(\alpha)=-\int_N
({{\mathscr{D}}}^{CN}\alpha_{ N})^\dag({{\mathscr{D}}}_{C}{}^L\eta_L).
\end{equation*}
 Using the identity
\begin{eqnarray*}
{ \mathscr D}_{AB} \lambda_{C}&=&{ \mathscr D}_{A(B}\lambda_{C)}
\,+\,{\mathscr D}_{A[B} \lambda_{C]}\nonumber\\
&=&{ \mathscr D}_{A(B}\lambda_{C)}
\,+\,\frac{1}{2}\epsilon_{BC}{ \mathscr D}_{AN} \lambda^ N,
\end{eqnarray*}
it may be checked that
\begin{equation*}
|a(\lambda, \alpha)|\le C\, || \lambda || \,|| \alpha || \,
\end{equation*}
for some constant $C$ and the linear functional $f$ is bounded.

To prove the coercivity of the bilinear form $a(\alpha,\lambda)$,
given $\psi_A$ is  supported in $N_{R_i}$, in general we
have
 \begin{eqnarray}
&&4\int_{N_{R_i}}|{{\mathscr D}}_A{}^N\psi_N|^2\nonumber\\
&=&\int_{N_{R_i}}\,\Big[\,(\mu\,|\psi|^2-j^an_a)+\frac{1}{8}|\psi|^{-2}\,|\dot h_{ab}|^2\nonumber\\
&&+2\,|D_{AB}\psi_C|^2-\frac{1}{2}|\psi|^{-2}\,|D_{(a}n_{b)}|^2\,
\nonumber\\
&+&\,\int_{S}\,\,-\sqrt 2\,\big(\psi^{\dag
A}\,\cancel\nabla_{A}{}^C\psi_C\,+\,\psi_{A}\,\cancel\nabla^{AC}\psi_C^{\dag}\,\big)\nonumber\\
&&-\int_{S}(K_{ab}\gamma^{ab}+p|\psi|^2)
\label{cor}
\end{eqnarray} where $|\psi|^2=\psi^{\dag A}\psi_A$ and $n_a$ is a shift vector to be specified.
As that  in the previous section, consider the partition $N=N_\epsilon\cup N/N_\epsilon$ and define a twisting of
$\psi_A$ in $N_\epsilon$ by
\begin{equation*}
\tilde\alpha_A=\eta\,\tilde\lambda_A+(1-\eta)\psi_A,
\end{equation*}
with $\tilde \lambda_A=|\psi|\tilde o_A$ where the flagpole of $\tilde o_A$ aligns with the future pointing null normal of $S$. A shift vector is chosen to be $n_a=\sqrt 2\,
\tilde\alpha_{(A}\tilde\alpha^{\dagger}{}_{B)}$. Subject further to the marginally trapped boundary condition imposed on $S$,  (\ref{cor}) then
becomes
\begin{eqnarray}
&&4\int_{N_{R_i}}|{{\mathscr D}}_A{}^N\psi_N|^2\nonumber\\
&=&\Big[\int_{N_{R_i}}\,\,(\mu\,|\psi|^2-j^an_a)\,+\,|\mathscr D_{AB}\psi_C|^2-C\epsilon\nonumber\\
&&+\,\int_{S}\,\,-\sqrt 2\,\big(\psi^{\dag
A}\,\cancel\nabla_{A}{}^C\psi_C\,+\,\psi_{A}\,\cancel\nabla^{AC}\psi_C^{\dag}\,\big)\Big].\label{cor1}
\end{eqnarray}
The APS boundary condition means that the inner boundary term in
(\ref{cor1}) is positive. Together with the dominant energy
condition and that $\epsilon$ is arbitrary,  (\ref{cor1}) may further be given as
\begin{eqnarray}
4\int_{N_{R_i}}|{{\mathscr D}}_A{}^N\psi_N|^2\ge
\int_{N_{R_i}}\,\,|{{\mathscr D}}_{AB}\psi_C|^2. \label{cor3}
\end{eqnarray}
With the index $i$ reinstated into the spinor field $\psi_A$ and from the definition of $ || \,||$, we may further infer from  (\ref{cor3}) that
\begin{equation*}
a( \psi_i, \psi_i)\ge\,C\,|| \psi_i ||^2.
\end{equation*}
for some constant $C$ independent of $i$.
 With $\eta_i$ in place of $\eta$ in (\ref{second-order}), a weak solution $\psi_{iA}$ exists for
(\ref{second-order}) . It may also be checked that $\psi_{iA}$ is uniformly
bounded in $H_{-}$, by passing to a subsequence if necessary
$\psi_{iA}$ converges weakly to some $\psi_A\in H_{-}$. Moreover,
it follows from the injectivity of the Sen-Witten operator  that $\psi_A$ is necessarily unique. Elliptic regularity then implies that
 $\psi_A$ is a strong, smooth solution to
(\ref{second-order}) with
the prescribed boundary conditions at the inner boundary and that near spatial infinity.
\section{Concluding Remarks}

The contribution of the present work lies in suggesting that a a spinor approach to the Penrose inequality is viable to a certain extent. The next  step towards
a complete proof of the Penrose inequality is to give an appropriate geometric
characterisation of an outermost trapped surface and see
whether spin geometry is capable of giving a lower bound of the norm of the Sen-Witten spinor field at the outermost trapped surface in terms of that of the
Schwarzschild metric.  In the course of development of the spinorial framework of the positive energy theorem, we have also uncovered certain geometric structures
of an initial data set underlying
the spinorial framework and might worth pursuing further. From a physical standpoint, the insights
we gain from the proof itself concerning the global structure
and geometry of an initial data set describing gravitational
collapse seem to be as valuable as the Penrose inequality
itself.

\vskip 20pt

\section*
{ Acknowledgement}

 The long term support by Professors Shing-Tung Yau and Lo Yang  to the research of general relativity in China
 through the Morningside Center of Mathematics
is crucial to the completion of this work.

%
%
%

\end{document}